%% file: emnlp2023.tex
\pdfoutput=1

\documentclass[11pt]{article}

\usepackage{EMNLP2023}

\usepackage{times}
\usepackage{latexsym}
\usepackage{amssymb, amsmath, amsfonts, graphicx, booktabs, adjustbox} 
\usepackage{listings} 
\usepackage[T1]{fontenc}

\usepackage[utf8]{inputenc}

\usepackage{microtype}

\usepackage{inconsolata}
\usepackage[capitalise,noabbrev,nameinlink]{cleveref}

\makeatletter
\newcommand{\todo}[2][red]{%
\textbf{\textcolor{#1}{#2}}
\@latex@warning{#2}{}{}}
\makeatother

\renewcommand{\todo}[2][]{}  

\newcommand{\jessy}[1]{\todo[orange]{J: #1}}

\newcommand{\methodname}{\textsc{Clarinet}}

\title{\textsc{Clarinet}: Augmenting Language Models to Ask Clarification Questions for Retrieval}

\author{Yizhou Chi\qquad Jessy Lin\qquad Kevin Lin\qquad Dan Klein\\ \{yizhouchi, jessy\_lin, k-lin, klein\}@berkeley.edu\\UC Berkeley \\}

\begin{document}
\maketitle
\input{sections/00-abstract}

\input{sections/10-intro}
\input{sections/20-related}

\input{sections/30-model}

\input{sections/40-experiments}

\input{sections/50-analysis}

\section{Discussion \& Conclusion}
Traditional heuristic approaches like EIG are often shown useful in the context of closed-form question selections where there is only a fixed number of vocabulary/answers and a limited amount of questions. 
The computation time increases dramatically if the answer becomes open-ended. Although the retrieval performance using KL as the question selection function is decent, it needs to estimate the resulting confidence distribution of every question in the candidate pool. It will be very costly and inefficient to run in a real-time interactive system.

We presented an interactive retrieval system that helps users retrieve books by asking open-ended clarification questions, finetuning a LLM to generate informative questions end-to-end.
Our \methodname{} model adopts an architecture that encodes the query, interactions, and passages separately so that the model could learn to ask questions that help identify the target more quickly with limited training data. 
We show that this approach can effectively distill the search over questions into the model, resulting in much cheaper inference while outperforming methods like EIG and KL that explicitly evaluate the usefulness of clarification questions at inference time.


\section*{Limitations}

The interactive retrieval system has not been evaluated by humans, so the retrieval performance could be different from a more realistic environment. Moreover, the system is capable of transferring to other domains or datasets. The neural question generator has shown decent retrieval performance, but its question generation preference is less interpretable than EIG and KL. Thus, it will be also helpful to develop a model that estimates EIG and KL more efficiently so that we can make more interpretable question selections.

\bibliography{anthology,custom}
\bibliographystyle{acl_natbib}

\appendix
\input{sections/99-appendix}

\twocolumn

\end{document}

%% file: sections/00-abstract.tex
\begin{abstract}
Users often make ambiguous requests that require clarification. We study the problem of asking clarification questions in an information retrieval setting, where systems often face ambiguous search queries and it is challenging to turn the uncertainty in the retrieval model into a natural language question.
We present \methodname{}, a system that asks informative clarification questions by choosing questions whose answers would maximize certainty in the correct candidate.
Our approach works by augmenting a large language model (LLM) to condition on a retrieval distribution, finetuning end-to-end to generate the question that would have maximized the rank of the true candidate at each turn.
When evaluated on a real-world retrieval dataset of users searching for books, our system outperforms traditional heuristics such as information gain on retrieval success by 17\% and vanilla prompted LLMs by 39\% relative.
\end{abstract}

%% file: sections/10-intro.tex
 \section{Introduction}

Natural language is a flexible interface for users to interact with systems, but language is inherently ambiguous and users themselves may not know what they want.
As a result, systems must handle underspecified queries.
We study this in an information retrieval setting, where search ambiguity is a well-studied challenge~\cite{keyvan2022approach}.

While modern large language models (LLMs) can ask coherent clarification questions, they do not always ask questions that elicit information specifically about what the model is uncertain about.
This is particularly challenging in the case of retrieval, where it is unclear how to integrate the conversational abilities of an LLM with the external database or the retrieval system that represents the uncertainty over search candidates.
In contrast, approaches that use principled information theoretic measures such as information gain \cite{oaksford1994rational, van2004utility, nelson2010experience,rothe2017question} or KL utility \cite{nelson2010experience, hawkins2017you} can use the retriever distribution to evaluate the right questions to ask and explicitly select questions that reduce uncertainty. However, these methods require expensive inference-time generation and evaluation of potential questions and have not yet been able to scale beyond toy settings.

In this work, we investigate whether we can learn to ask good questions simply by distilling the search over good questions into an end-to-end model.
We present \methodname{}, a framework for learning to ask clarification questions for information retrieval. Our approach works by augmenting a language model to condition on the retriever distribution and then finetuning the system end-to-end to generate informative questions.
We select informative questions by simulating interactions with a prompted LLM that acts as a user proxy, training only on questions that would have significantly increased the confidence in the true item if answered.
In contrast to heuristic-based methods, \methodname{} distills the expensive, explicit search over questions at inference time into the model.
Then, given user responses to clarification questions, we summarize the interaction history into a \emph{language posterior}, a single natural language query describing what the system knows about the user's desired item. We use this query to re-rank candidate items from the database.

\begin{figure*}[ht!]
    \includegraphics[width=\textwidth]{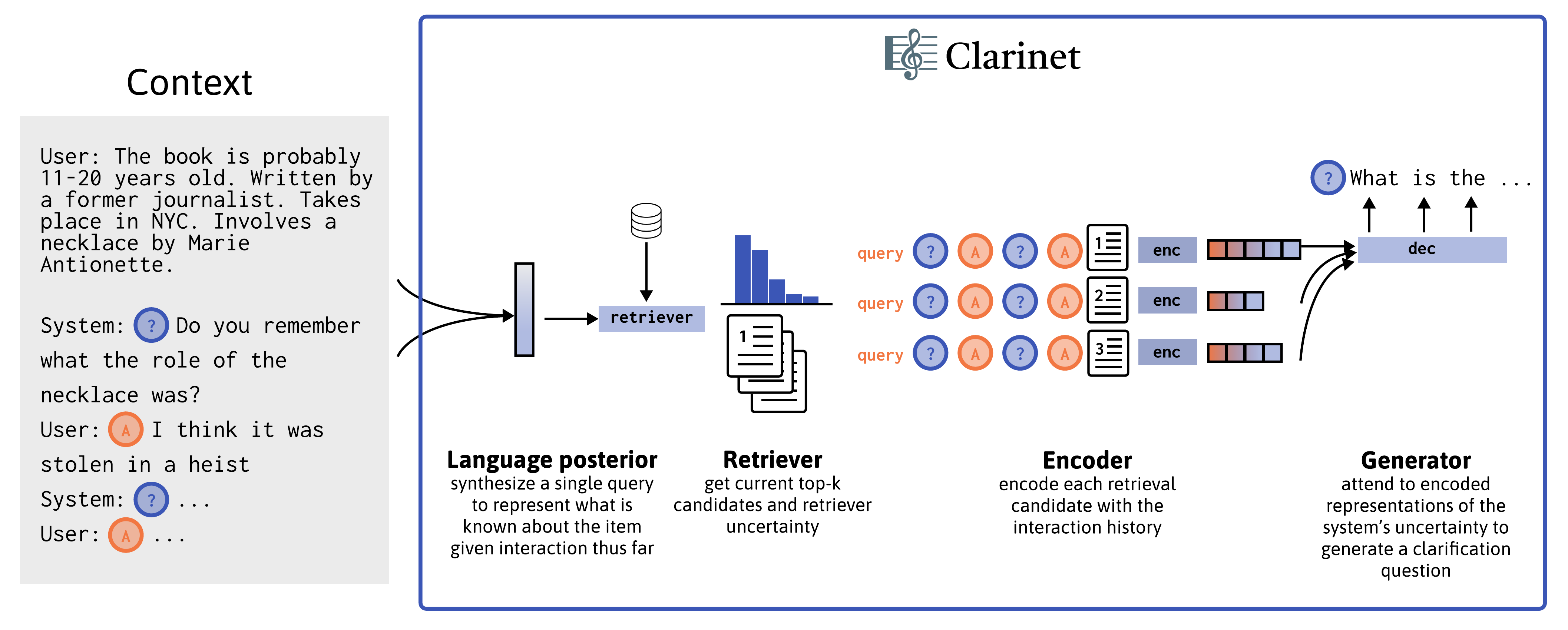}
    \caption{The system takes in the user's initial query and past interactions and summarizes the content before feeding them into the retriever. The retriever computes the corresponding score for each candidate and outputs a confidence distribution. The system then encodes each passage concatenated with the context separately, and these embeddings will be supplied to the decoder to generate the next clarification question.\todo{TODO}}
    \label{fig:clarinet}
\end{figure*}

We evaluate our approach on a real-world dataset of users asking for help on an online forum (Goodreads) to find books they vaguely recall from a database of thousands of items, e.g., ``written by a former journalist, takes place in NYC, involves a necklace''~\citep{bhargav2022s}. We evaluate systems interactively against a simulated user, which we implement as a prompted LLM that can answer clarification questions with oracle access to the true item. 
Compared to a purely dialogue-based approach that generates clarification questions prompted only with the dialogue history, our system asks better questions to achieve a 39\% relative gain in top-1 retrieval accuracy after ten interactions with the simulated user.
Our model also outperforms heuristic-based approaches that select a clarification question from a candidate pool with information gain or KL utility by 17\% relative on top-1 retrieval accuracy, while being much simpler and cheaper at inference time.


%% file: sections/20-related.tex
\section{Related Work}

Interactive NLP systems have used clarification questions as way to gather more information from users in settings such as classifying \cite{yu2019interactive}, conversational question answering \cite{rao2018learning}, and visually grounded 20 questions games \cite{white2021open}. To select informative questions, much of the work draws heuristics that maximize expected information gain from simulated answers.


Some of the works shift the clarification question generation and selection from using rule-based and heuristic methods to reinforcement learning. \cite{rao2019answer} extended their utility used in \cite{rao2018learning} in a reinforcement setting to generate useful and relevant questions. Meanwhile, \cite{pyatkin2022reinforced} presented an interactive system that asks relevant and informative clarification questions to learn salient contexts of a social or moral situation. Their approach to question generation utilized reinforcement learning, aiming to optimize the generation of questions that elicit responses containing morally relevant contexts. 

In a more practical scenario, \cite{zamani2020generating} suggests both supervised and reinforcement learning models to generate clarifying questions that aid users in formulating search queries in search engines. Additionally, they explore techniques for generating potential answers for each clarifying question, allowing users to conveniently choose from a predefined set of answers. However, if the users don't know what they are looking for, the predefined answers may not be very helpful.

\todo{see related work of nelson 2010: bayesian optimal experiment design framework}





%% file: sections/30-model.tex
\section{Method}

\paragraph{Dataset} We train and evaluate our system in a ``tip of the tongue'' (TOT) search setting, where users are searching for items they vaguely remember but for which they cannot formulate precise queries. To train and evaluate clarification question generation, we use the dataset from \citet{bhargav2022s}, which contains multi-turn interactions from community forums where users post queries searching for books where they were unable to find using conventional search engines. We filter the data specifically about book queries, resulting in 784 interactions, which we split into 156 for training and 628 for evaluation. The retrieval database consists of a corpus of 1877 documents.

\paragraph{Retriever}
In general, \methodname{} is agnostic to the choice of retrieval model. In our experiments, we use Dense Passage Retriever (DPR)~\cite{karpukhin2020dense} as the retriever over the book database. 
We finetune the retriever for this task with the larger scale \textit{WhatsThatBook} dataset from \citet{lin2023decomposing}, which contains 11,552 initial query-document pairs for TOT book retrieval.
We use dot-product similarity between the DPR representation of the book information (book title, author, metadata, synopsis) and the DPR representation of a query synthesized by our \methodname{} model (as described in~\cref{sec:method-clarinet}) to rank candidate books.







\subsection{Asking Clarification Questions with \methodname{}}
\label{sec:method-clarinet}

The retrieval dataset itself consists of initial (underspecified) queries from real users, and the true item they were searching for.
To train a \methodname{} model, we transform a retrieval dataset into an interaction dataset by synthesizing a series of dialogue interactions with a user simulator model. Each interaction is seeded with a real user query from the dataset; we then generate a series of clarification questions and answers from the user simulator model.
At each turn, we generate a pool of \emph{candidate questions} and filter for the most informative questions for finding the user's desired items---i.e., the questions whose answers would increase the rank of the true candidate the most.
The clarification dataset thus consists of \texttt{(interaction history $(q_0, a_0, \ldots, q_{t-1}, a_{t-1})$, $q_t$)} for each informative question $q_t$ out of the candidate pool.
We finetune a LLM to generate these questions $q_t$ conditioned on the interaction history and retrieval distribution.\jessy{still kind of unclear}
For the TOT dataset, we run five 10-turn games for each initial user query.

At inference-time, we directly sample from the model to generate questions to ask the user. We update the retrieval distribution after each user response by condensing the interaction history thus far into a \emph{language posterior}, a single retrieval query that we use to re-retrieve results from the database.

\paragraph{Candidate Question Generation} At each turn, we generate a pool of 50 candidate clarification questions by prompting GPT-3 (\texttt{gpt-3.5-turbo-0613}) with the user's initial query (dataset queries from real users), the dialogue history of clarification questions and answers, and top three items in the retriever distribution. We generate 20 candidate clarification questions for each turn by sampling with temperature 0.8.

\paragraph{Question Selection} For each candidate question, we simulate a response with a user simulator model. We implement the user simulator as a GPT-3 model that has access to the true item and its description, and is prompted to answer questions about the item vaguely. Refer to~\cref{sec:appendix-prompts} for the prompt.

For each initial training query, make a run where the question at each turn is uniformly sampled from the candidate pool generated by GPT-3, and the retrieval distribution is updated through explicit posterior. At the end of each turn, calculate and record the ranking of the target book.

In contrast to methods like EIG that operate at inference time, we can directly train on the question that places the target's rank high or improves the rank the most.
Only the questions that help the system to rank the target book as top 10, or help increase the rank by 10 will be used in the training. Users' initial queries, past interactions, and the information of top book candidates (book candidates with retrieval probabilities that add up to 50\% [no more than 3 books]) will be served as the model input. 

\begin{figure*}[h]
  \centering
  \includegraphics[width=1\textwidth]{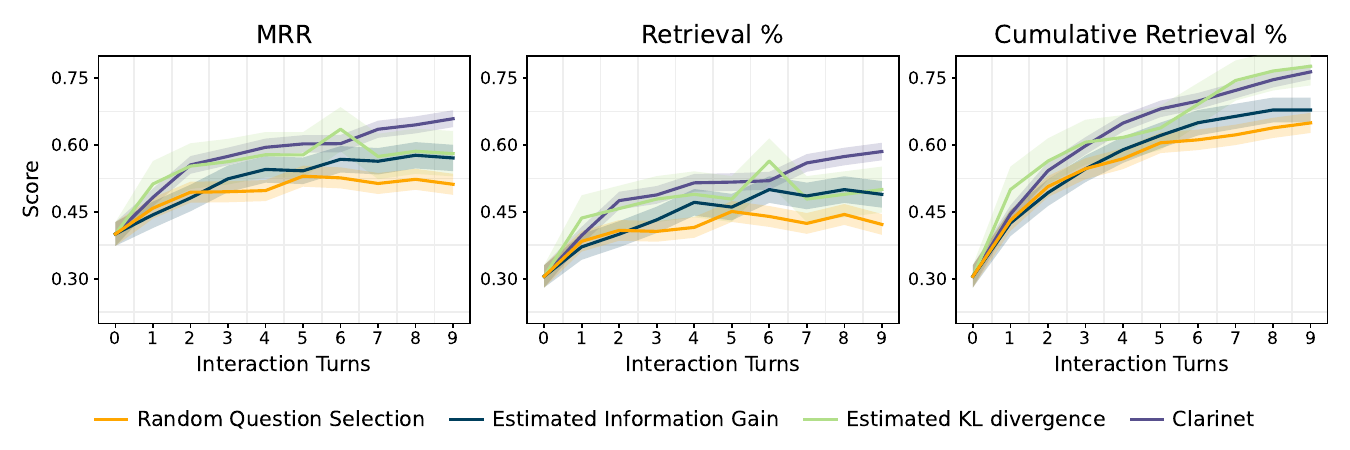}
  \caption{The retrieval performance for random question selection, EIG, KL, and our model with SEM error bars}
  \label{main}
\end{figure*}
\paragraph{Language Posterior} To produce an updated belief distribution over the retrieval candidates after each turn, we need to integrate all the information that the user has provided thus far. We prompt another GPT-3 model to synthesize a description of the retrieval candidate given the interaction history thus far, which can be thought of as a posterior belief over the true item, represented in language. We then use this \emph{language posterior} as the search query to produce a new candidate distribution, instead of the initial query provided by the user.

\paragraph{Training} We train Flan-T5-base \cite{chung2022scaling} to generate the selected clarification questions.
The model takes in the user's initial query, the interaction history of questions and answers, and the information of the top 50\% (no more than 3) confident retrieval candidates to generate the next clarification question.
For the books dataset, the information for each retrieval candidate includes the book title, author, published dates, and description. We additionally include the current ranking and retrieval probability, concatenating all the information as text to form the full context for a book.
We use a Fusion-in-Decoder architecture (FiD) \cite{izacard2020leveraging}, where we concatenate the information for each retrieval candidate with the initial query and interaction history and feed it into the encoder independently. 
The encoded candidates are then concatenated, which the decoder then attends to in order to generate clarification questions.

We train the model for 10k gradient steps using the Adam optimizer with a learning rate of $10^{-4}$, batch size 8, and dropout 0.1. We evaluate the models at intervals of 500 steps and select the checkpoint with the best BLEU score on a held-out validation.

%% file: sections/40-experiments.tex
\begin{figure*}[h]
  \centering
  \includegraphics[width=1\textwidth]{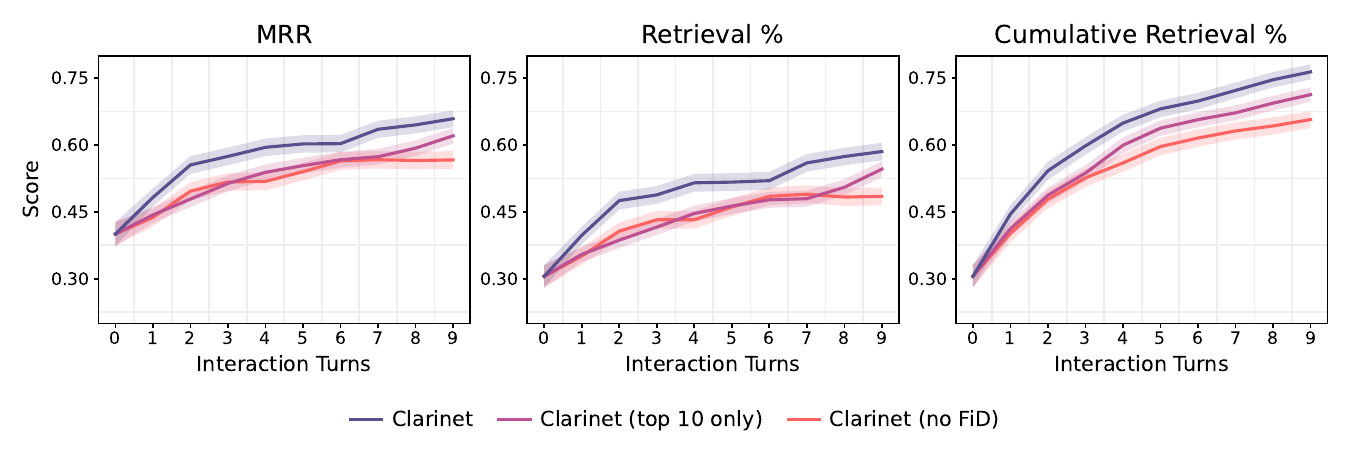}
  \caption{The retrieval performance for model variants.}
  \label{variants}
\end{figure*}
\section{Experiments}

In our experiments, we aim to answer the following questions:
\begin{itemize}
    \item How does our method compare to well-studied methods for asking both quantitatively (how effectively do our system's questions help us narrow down the user's item) and qualitatively (how do our system's questions differ)?
    \item How important is conditioning on information from the retriever vs. purely dialogue- or prompting-based approaches that reason purely in text and generate questions conditioned only on the dialogue history?
    \item What types of questions are helpful in increasing confidence in the target and ultimately achieving successful retrieval?

\end{itemize}
In this section, we present the system's empirical performance by evaluating it in a book retrieval setting. In one retrieval game, the system will be given an initial query that vaguely describes the book that the user is looking for. The system will then be allowed to ask 9 clarification questions to identify the book (a total of 10 turns including the initial query).

We run experiments with a simulated user, which we implement as a GPT-3 model prompted with oracle information about the target book, removing the title so that the user simulator cannot output the name of the target item outright. We prompt the user simulator to mimic a user is searching for the book with a vague memory about the content.

We measure top-1 retrieval accuracy since there is one correct retrieval item. We compare model performance on cumulative retrieval success, where an interaction up to turn $t$ is counted as successful if the correct item is retrieved at any of turn $0,\ldots,t$ and non-cumulative retrieval success, where the interaction is successful only if the correct item is the top candidate at turn $t$.
We additionally show trends for mean reciprocal rank (MRR), the average reciprocal rank of the correct item, to contextualize model performance beyond top-1.

\subsection{Baselines}

We compare \methodname{} to a purely dialogue-based approach that randomly generates a question conditioned only on the dialogue history, without access to the retriever distribution. This model simply randomly samples a question from the pool of candidate questions generated by GPT-3, prompted with the initial query and dialogue history.

Additionally, we implement two information theoretic approaches to clarification question generation frequently used in prior work \cite{rao2018learning, yu2019interactive, white2021open}. These approaches typically generate a pool of candidate questions at each turn, using different notions of question usefulness to select the best question to ask.
In contrast to \methodname{}, because these approaches explicitly evaluate candidate questions, they are much more expensive at inference time.
We implement question selection based on expected information gain and Kullback-Liebler (KL) divergence, which we describe in more detail below.

\subsubsection{Expected Information Gain (EIG)}
Expected information gain (EIG) selects clarification questions that are most likely to yield the most information, in expectation over potential user responses. Formally, at turn $t$ we want to choose the question $q_t$ that optimizes:
\[
    q_t = \arg\max_q \mathbb{E}_{p_{t-1}(y \mid q)} [I(Y; a \mid q)]
\]
where $p_{t-1}(y)$ is the model's current posterior belief over the true item $y$ with the information
up to turn $t-1$.
$I(Y;a \mid q)$ is the information gain (or equivalently, entropy reduction) in the candidate distribution $Y$ from observing answer $a$, given that we asked $q$:
\begin{align*}
    I(Y;a\mid q) &= H(Y \mid q) - H(Y \mid a, q) \\
                 &= H(A \mid q) - H(A \mid y, q)
\end{align*}
Thus, to find the most informative question, the optimization problem can be simplified to:
\begin{align*}
    q_t = \arg\max_q H (A \mid q) - \mathbb{E}_{p_{t-1}} [H(A \mid q, y)]
\end{align*}
where the answer distribution is obtained by marginalizing over the current belief distribution $p(y)$ (subscript $t$ omitted):
\begin{align*}
    p(a \mid q) = \sum_{y} p(a \mid q, y) p(y) 
\end{align*}

To compute the answer distribution $p_t(a \mid q, y)$, we answer every question $q$ for every candidate $y$ with a pre-trained Flan-T5-base model. The answer probabilities $p(a \mid q, y)$ are estimated by softmaxing over the summation of logits of tokens across every generation step.
\begin{figure*}[h]
  \centering
  \includegraphics[width=1\textwidth]{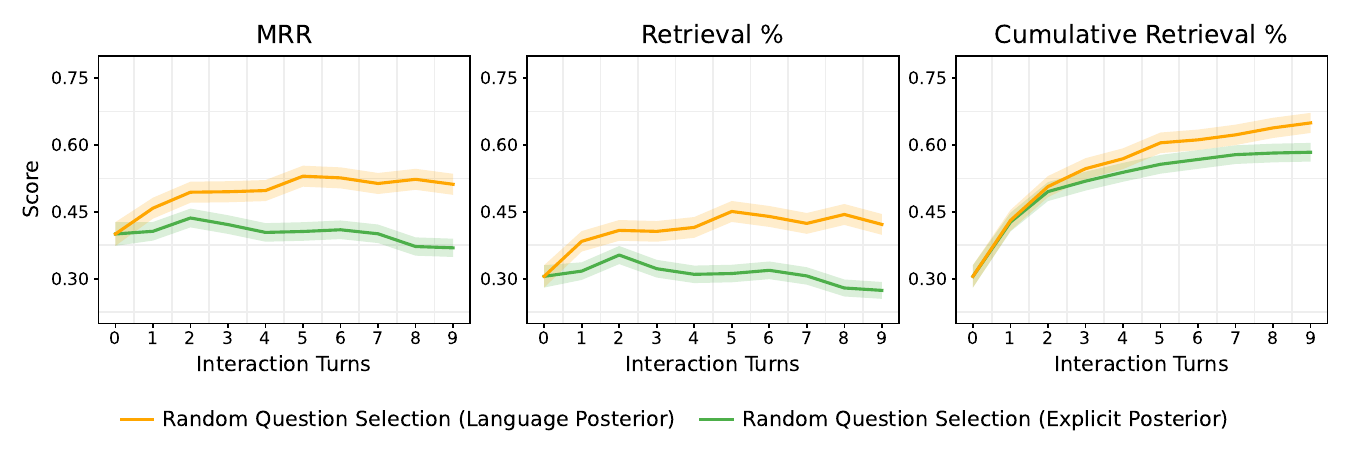}
  \caption{Language Posterior vs Explicit Posterior}
  \label{pu vs rs}
\end{figure*}
\subsubsection{KL Divergence}
The previous heuristic function helps pick the question candidate with the highest expected information gain. However, when there are only a few books with high confidence, the question selector using EIG will only try to select a question that differentiates between the top candidates. As a result, the selected questions in the subsequent turns can become very similar. Therefore, we'd also like a heuristic function that selects the question that is likely to \emph{change} the current belief distribution. 

\begin{align*}
    q_t = \arg\max_q p_t(y | q, a) \log\frac{p_t(y | q, a)}{p_{t-1}(y)}
\end{align*}


To estimate the posterior after observing an answer to a question $q$, $p_t(y \mid q, a)$, we answer every question $q_i$ for every candidate item with a pretrained Flan-T5-base model, like the EIG baseline.
\jessy{We then approximate the single-turn likelihood $p(y \mid q, a)$ by making the simplifying assumption that we will see answers from this set of candidate answer}

$p(y_i \mid q_i, a_i)$ is estimated by computing the cosine similarity between the answers calculated from different (question, book) pair.

\[
    P(y_{i} | q_t^{i}, a_t^{i}) = \frac{1}{n} \sum\limits_{j=1}^{n} \frac{E(a_{i, ref}) \cdot E(a_{j, ref})}{\|E(a_{i, ref})\|\|E(a_{j, ref})\|}
\]

where $ E(a) $ is the language embedding of the referenced answer.

\subsection{Results}

In \cref{main}, we show that \methodname{} achieves higher retrieval success than the dialogue-only baseline that randomly selects questions given only the dialogue context, as well as outperforming the information theoretic approaches. The performance of random question selection plateaus after a certain number of turns\todo{say a bit more about the other methods here}.

%% file: sections/50-analysis.tex
\section{Analysis}
\begin{figure*}[h]
  \centering
  \includegraphics[width=1\textwidth]{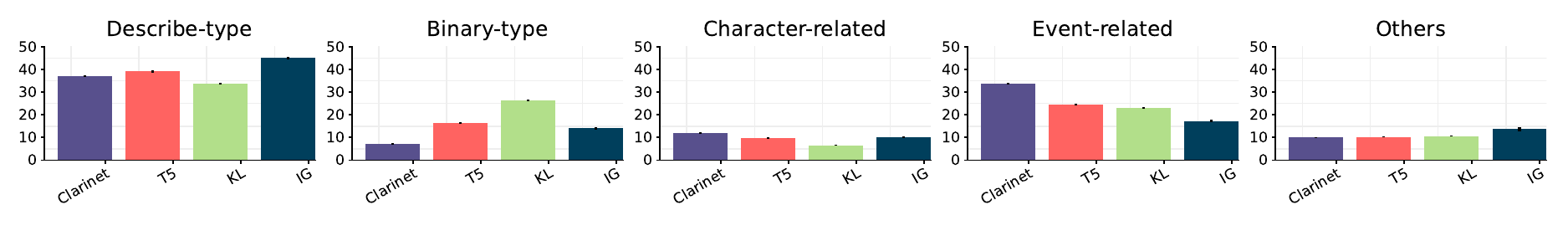}
  \caption{For every 100 questions (roughly 10 retrieval games) that each objective  function selects, the number of \textbf{describe-type} questions, \textbf{binary} questions, \textbf{character-related} questions, \textbf{event-related} questions, and others; the questions are labeled by GPT3.5-turbo-0613}
  \label{kind}
\end{figure*}

\subsection{Qualitative Analysis}

To understand how the \emph{kinds} of clarification questions that models ask may differ, we label every 100 questions generated by each model with GPT-3, prompting it to label each question as one of four categories or ``other.''
According to figure \ref{kind}, the clarification questions generated by \methodname{} are shown qualitatively different than the other methods. Specifically, it does not generate many binary questions as it might learn that the binary questions are not providing as much information. Additionally, \methodname{} generates more questions that are specific to characters or events than the other methods. This behavior could potentially show that events and scenes are one of the determining factors that distinguish between books.

As \cref{kind} suggests, there are also notable differences between the information gain and KL divergence heuristics for question selection. The KL divergence approach tends to ask more binary questions and fewer describe-type questions than EIG question selector. The behavior is reasonable as KL heuristic likes to select questions that could potentially shift the distribution a lot. A binary (yes-no) question is more likely to change the distribution than an open-ended question as the answer to the binary question is polarized. In general, different question-asking behaviors may be effective in different contexts; while KL performed better on retrieval accuracy in our settings, open-ended questions provide more information than a binary question in theory. The type of behavior that is preferable may depend on how the user responds to queries, the robustness of the retrieval system, among other qualities.


\subsection{Additional Analysis}

Next, we investigate the effect of the language posterior, FiD architecture, and gain in rank for train-time question selection.

\paragraph{Condensing the interaction into a language posterior drastically improves over updating an explicit posterior.}
We evaluate the effect of summarizing the interaction so far into language posterior, compared to maintaining a explicit posterior belief distribution that is updated at each turn with Bayes' Rule, i.e. $p_t(y \mid q_t, a_t) \propto p(q_t, a_t \mid y) p_{t-1}(y)$. We compare random question selection with each of these posterior belief representations. As shown in \cref{pu vs rs}, the language posterior has significantly higher retrieval performance. We observe that the MRR of the true item \emph{decreases} with more questions when using an explicit posterior, despite the fact that more information should theoretically always help the model improve its belief about the true item. Qualitatively, we observe that the explicit posterior fails because it is less robust to errors, which may compound at each point in the system or over course of several turns. For example, if the human or user simulator provides inaccurate information to the system or if the retrieval model fails to interpret a particular user response, a single turn can have a large impact on the belief distribution that is hard to correct.

Additionally, single turn responses to clarification questions are more out-of-distribution for the retriever, whereas the language posterior synthesizes the interaction history into something that looks like a hypothetical query.

One downside of the language posterior that we observed qualitatively was that initial queries, which were often much longer than responses to clarification questions, were over-represented in the language posterior. Future work may improve gains in the early turns by emphasizing or upweighting the question responses when retrieving in the early turns.

\paragraph{The fusion-in-decoder architecture and selecting questions by gain in rank is important for effective clarification question generation.}

We also compare the performance of another variant of our model as well as a fine-tuned FLAN-T5-base model. In figure \ref{variants} the variant annotated with "top 10" is trained with only the questions that help the retriever rank the target book as top 10, as opposed to the "delta" model that is trained with the questions that either help the retriever ranks the target as top 10 or increase the target's rank by 10. The better performance of the "delta" model suggests that the questions that help increase the target's rank are as important as questions that give the target a high absolute rank. The finetuned t5 model is trained with the same dataset and parameters as the "delta" model except that the initial query, interactions, and top book information will be put into a single text string and fed into the encoder.

The comparison between our model with FiD architecture and the fine-tuned t5 suggests that it is very hard for the model to explicitly represent uncertainty given text descriptions with small data. Instead, encoding the passages separately would help the model ask a more useful question that could more efficiently identify the target.

%% file: sections/99-appendix.tex
\clearpage
\onecolumn
\section{Retrieval Results}
\label{sec:appendix-results}
\begin{table*}[ht!]
\adjustbox{max width=\textwidth}{%
    \centering
    \begin{tabular}{lrrrr}
    \toprule
               Methods &    MRR &  Retrieval Rate &  Cumulat. Retrieval Rate &  $\Delta$ Cumulat. Retrieval Rate \\
    \midrule
 Linear Search & 0.0025 &       0.0013 &                  0.0128 &                              0.0115 \\
Random (language posterior) & 0.5118 &       0.4219 &                  0.6496 &                              0.3442 \\
Random (explicit posterior) & 0.3694 &       0.2739 &                  0.5838 &                              0.2784 \\
                   EIG & 0.5709 &       0.4893 &                  0.6786 &                              0.3732 \\
                    KL & 0.5806 &       0.5000 &                  0.7766 &                              0.4712 \\
     Clarinet (without FiD) & 0.5665 &       0.4848 &                  0.6571 &                              0.3517 \\
    Clarinet (top 10) & 0.6206 &       0.5459 &                  0.7130 &                              0.4076 \\
     Clarinet (delta) & 0.6590 &       0.5853 &                  0.7640 &                              0.4586 \\
    \bottomrule
    \end{tabular}
}
    \caption{Mean retrieval performance after 10 turns (9 questions being asked). $\Delta$ cumulative retrieval \% represents the increase in retrieval rate between the first turn and the final turn}
\end{table*}

\clearpage
\onecolumn
\section{Prompts}

\label{sec:appendix-prompts}
\textbf{\texttt{Question Generation One-shot Example} (t = 0)}
\par\noindent\rule{\columnwidth}{0.4pt}
\begin{lstlisting}[breaklines=true]
Book information:
author: Kevin Henkes
description: Sometimes life can change in an instant. Martha Boyle and Olive Barstow could have been friends, but they weren't. Weeks after a tragic accident, all that is left are eerie connections between the two girls, former classmates who both kept the same secret without knowing it. Now, even while on vacation at the ocean, Martha can't stop thinking about Olive. Things only get more complicated when Martha begins to like Jimmy Manning, a neighbor boy she used to despise. What is going on? Can life for Martha be the same ever again?
date: Published
        April 26th 2005
         by Greenwillow Books

              (first published January 1st 2001)
genres: {'Young Adult': 59, 'Fiction': 152, 'Realistic Fiction': 148, 'Childrens': 45, 'Literature': 37, 'Death': 33, 'Contemporary': 31}

Query:
*Spoilers ahead*The protagonist was 12/13 years old but the book was somewhat mature for that age, not a children's book. I think there were 5/6 brothers that she hung out with all summer. She started the summer with a crush on Tate but spent more time with his older brother Jimmy. He had her help him with the movie he was making, then kissed her and filmed it to win a bet with one of his other brothers. At the end of summer Tate tells her he liked her all along. I believe part of the plot was that her friend (maybe from Madison, WI?) had never been to the ocean so she brought her home a bottle/jar of water. I read this for a book report in 2012 and i am so frustrated that I can't remember or find it anywhere. Any help is appreciated!

Based on the query, ask one more clarification question to learn more information about the book that the person is looking for.
In the question, you must not mention any of the book's name or specific characters.
Questions: 
Is there a tragic accident in the story?
Did the girl hate Jimmy at the beginning?
\end{lstlisting}
\par\noindent\rule{\columnwidth}{0.4pt}

\textbf{\texttt{Question Generation} Prompt (t = 0)}
\par\noindent\rule{\columnwidth}{0.4pt}
\begin{lstlisting}[breaklines=true]
Book Information:
{} 
Query:
{} 
Based on the query, ask one more clarification question to learn more information about the book that the person is looking for.
In the question, you must not mention any of the book's name or specific characters.
Clarification Questions: 
\end{lstlisting}
\par\noindent\rule{\columnwidth}{0.4pt}

\textbf{\texttt{Question Generation One-shot example} Prompt (t > 0)}
\par\noindent\rule{\columnwidth}{0.4pt}
\UseRawInputEncoding
\begin{lstlisting}[breaklines=true]
Query:
I read this book in the 3rd grade, which was 1996 but the book was probably written in the 80s. I don't remember much except what the headline says. Basically these two kids (I want to say a brother and sister, around 10 years old or so) are wandering around and happen upon all these old Victorian homes that have been abandoned except one. An old woman still occupies it and at first they are scared but she then tells them about the history of the area.These are the top candidate books that are likely to be the goal of the search.
Top candidate books:
[Book 0]
author: M.L. Forman
description: Do you have the courage, the wits, and the skill to claim a dragon's hoard? If so, apply within... The sign is small, tucked into the corner of Mr. Clutter’s bookshop window: “Adventurers Wanted. Apply Within.” No one but fifteen-year-old Alex Taylor even seems to notice it is there. And for Alex, who has wished for a change in his life, it is an irresistible invitation.Upon Do you have the courage, the wits, and the skill to claim a dragon's hoard? If so, apply within ...The sign is small, tucked into the corner of Mr. Clutter’s bookshop window: “Adventurers Wanted. Apply Within.” No one but fifteen-year-old Alex Taylor even seems to notice it is there. And for Alex, who has wished for a change in his life, it is an irresistible invitation.Upon entering Mr. Clutter's shop, Alex is swept away on an incredible adventure to a faraway land filled with heroic warriors, mysterious elves, and hard-working dwarves. Alex becomes the eigth man in a band of adventurers seeking the lair of Slathbog the Red - and evil dragon with a legendary treasure. Along the way, Alex and his new friends must battle dangerous trolls and bandits, face undead wraiths, and seek the wisdom of the Oracle in her White Tower. Alex’s adventure takes him to distant and exotic lands where he learns about courage, integrity, honor, and, most importantly, friendship.Slathbog’s Gold is the first book in an exciting new YA epic fantasy series and heralds the arrival of a major new talent in the genre.

date: Published
        February 1st 2009
         by Shadow Mountain
genres: {'Fantasy': 44, 'Young Adult': 9, 'Adventure': 75, 'Fiction': 40, 'Childrens': 12}
[Book 1]
author: Neal Shusterman
description: Nick and Allie don't survive the car accident...  ...but their souls don't exactly get where they're supposed to get either. Instead, they're caught halfway between life and death, in a sort of limbo known as Everlost: a shadow of the living world, filled with all the things and places that no longer exist. It's a magical, yet dangerous place where bands of lost children r Nick and Allie don't survive the car accident...  ...but their souls don't exactly get where they're supposed to get either. Instead, they're caught halfway between life and death, in a sort of limbo known as Everlost: a shadow of the living world, filled with all the things and places that no longer exist. It's a magical, yet dangerous place where bands of lost children run wild and anyone who stands in the same place too long sinks to the center of the Earth.  When they find Mary, the self-proclaimed queen of lost kids, Nick feels like he he's found a home, but Allie isn't satisfied spending eternity between worlds. Against all warnings, Allie begins learning the "Criminal Art" of haunting, and ventures into dangerous territory, where a monster called the McGill threatens all the souls of Everlost.  In this imaginative novel, Neal Shusterman explores questions of life, death, and what just might lie in between. ...more
date: Published
        October 1st 2006
         by Simon  Schuster Books for Young Readers
genres: {'Fantasy': 124, 'Young Adult': 66, 'Fiction': 208, 'Paranormal': 114, 'Science Fiction': 96, 'Adventure': 88, 'Death': 65}
[Book 2]
author: Kelly Link
description: A lonely boy in bungalow 6 is taunted by the other kids on camp. Then he meets a monster in the snow, a monster with a sense of humour. He is the only one to talk to it, but only after it has eaten the rest of bungalow 6. Dorn misses his soccer match when his father kidnaps him to be held in quarantine under machine-gun guard in Costa Rica-while they wait for the aliens to A lonely boy in bungalow 6 is taunted by the other kids on camp. Then he meets a monster in the snow, a monster with a sense of humour. He is the only one to talk to it, but only after it has eaten the rest of bungalow 6. Dorn misses his soccer match when his father kidnaps him to be held in quarantine under machine-gun guard in Costa Rica-while they wait for the aliens to return. Zilla, a ghost , is not greedy. She does not bleed her clients dry; she milks them. After all, there is only so much blood a grown woman and a smallish girl have to spare. The world is full of things and nobody ever sees them! Nobody except for you and me. And you will see the things that nobody sees...inside Pretty Monsters the companion volume to The Wrong Grave, from the highly acclaimed sorceress of fantasy stories, Kelly Link. Librarian note: Despite having the same title, this edition contains only half of stories from the original "Pretty Monsters", with the other half available in "The Wrong Grave".  ...more
date: Published
        May 31st 2010
         by Text Publishing

              (first published May 30th 2010)
genres: {'Short Stories': 13, 'Fantasy': 12, 'Fiction': 7, 'Young Adult': 6, 'Horror': 5, 'Science Fiction': 3, 'Anthologies': 3, 'Adult': 2}
Interactions:
Q: Can you recall any notable events or conflicts that happened in the story?
A: In the story, the main characters stumble upon a mysterious message on a boulder and explore a swamp that used to be a beautiful lake. They also come across a ghost town and encounter two people who have never left.
Q: Can you describe the characters' personality in terms of specific location or location?
A: Yes, there are characters in this book whose personalities are influenced by a specific location or location.

Based on the interactions, ask more clarification questions to learn more information about the book that the person is looking for.
In the question, you must not mention any of the book's name or specific characters.
Try to ask diverse clarification questions that are different from the questions asked in the interactions. The questions can either be open-ended or binary.

Clarification Questions:
Are there dragons in the book?
Is there any sort of soul world?
Where does the story take place?
Can you describe the magical creatures or monsters if it's mentioned in the book?
\end{lstlisting}
\par\noindent\rule{\columnwidth}{0.4pt}

\textbf{\texttt{Question Generation} Prompt (t > 0)}
\par\noindent\rule{\columnwidth}{0.4pt}
\begin{lstlisting}[breaklines=true]
Query:
{} 
These are the top candidate books that are likely to be the goal of the search.
Top candidate books:
{} 
Interactions:
{} 
Based on the interactions, ask more clarification questions to learn more information about the book that the person is looking for.
In the question, you must not mention any of the book's name or specific characters.
Try to ask diverse clarification questions that are different from the questions asked in the interactions. The questions can either be open-ended or binary. 

Clarification Questions:
\end{lstlisting}
\par\noindent\rule{\columnwidth}{0.4pt}

\textbf{\texttt{Answer simulator} Prompt}
\par\noindent\rule{\columnwidth}{0.4pt}
\begin{lstlisting}[breaklines=true]
Book Information:
{} 
Suppose you have read this book, but you only have a very vague memory of the book and are looking for this book. 
Answer the question vaguely. You must not mention any of the book's name or specific characters. Do not include too specific information.
{}
\end{lstlisting}
\par\noindent\rule{\columnwidth}{0.4pt}

\textbf{\texttt{Response summarizer} Prompt}
\par\noindent\rule{\columnwidth}{0.4pt}
\begin{lstlisting}[breaklines=true]
Condense the following descriptions into one statement:
{}
\end{lstlisting}
\par\noindent\rule{\columnwidth}{0.4pt}